\documentclass[aps, prl, superscriptaddress,reprint,amsmath,amssymb]{revtex4-2}
\usepackage[]{hyperref}
\hypersetup{
    colorlinks=true,
    linkcolor=blue,
    filecolor=magenta,      
    urlcolor=cyan,
    citecolor = blue
    }

\usepackage{graphicx}
\usepackage{dcolumn}
\usepackage{bm}
\usepackage{xcolor}
\usepackage{comment}

\newcommand{\affilPhotonics}{Photonics Laboratory, ETH Zürich, 8093 Zürich, Switzerland}
\newcommand{\affilQC}{Quantum Center, ETH Zürich, 8093 Zürich, Switzerland}
\newcommand{\affilMerkt}{Laboratory of Physical Chemistry, ETH Z\"urich, 8093 Z\"urich, Switzerland}
\newcommand{\affilDelft}{Kavli Institute of Nanoscience, Department of Quantum Nanoscience,
TU Delft, 2628CJ Delft, The Netherlands}

\begin{document}

\title{Single-atom detection with a quantum-controlled mechanical oscillator}

\author{Maxime Perdriat}
\affiliation{\affilPhotonics}
\affiliation{\affilQC}
\author{Maciej Dziewiecki}
\affiliation{\affilPhotonics}
\affiliation{\affilQC}
\author{Josef-Anton Agner}
\affiliation{\affilMerkt}
\affiliation{\affilQC}
\author{Massimiliano Rossi}
\altaffiliation[Present address: ]{\affilDelft}
\affiliation{\affilPhotonics}
\affiliation{\affilQC}
\author{Frédéric Merkt}
\affiliation{\affilMerkt}
\affiliation{\affilQC}
\author{Martin Frimmer}
\affiliation{\affilPhotonics}
\affiliation{\affilQC}
\author{Lukas Novotny}
\affiliation{\affilPhotonics}
\affiliation{\affilQC}


\begin{abstract}
Using supersonic expansion, we create a time-gated, directional beam of Xe atoms with a narrow momentum distribution and detect their collisions with a levitated nanosphere cooled to its quantum ground state. We observe individual momentum kicks below 50 keV/c with $96\%$ confidence and distinguish directional momentum transfer from the atomic beam against the background of thermal collisions.
Our experiments constitute a first step toward the exploration of distance-dependent short-range interactions between atoms and levitated systems, as well as toward the use of levitated platforms for impulsive force sensing in previously unexplored parameter regimes.
\end{abstract} 

\maketitle

\paragraph{Introduction.}
{\color{black} Owing to their low mass, high degree of isolation and controllability under ultra-high vacuum conditions, levitated nanospheres have recently emerged as a promising platform for the sensing of static, oscillating and impulsive forces~\cite{gonzalez2021levitodynamics,jin_towards_2024}. Its high force sensitivity is being explored for applications in fundamental physics, such as the preparation of macroscopic quantum states~\cite{romero-isart_quantum_2011} or the search for dark matter~\cite{moore2021searching}, but also for technological applications, such as 
magnetic and electrical field sensing~\cite{hempston_force_2017,timberlake_acceleration_2019,shi_mobile_2025,ahrens_levitated_2025}
 pressure sensing~\cite{liu_nanoscale_2024}, charge and
mass sensing~\cite{moore_search_2014,frimmer_controlling_2017,ricci_accurate_2019}, as well as gravitational
and inertial force sensing~\cite{fuchs_measuring_2024,hebestreit_sensing_2018,monteiro2020search,zeng_optically_2024,zielinska_long-axis_2024}. 

In impulsive force sensing, the interaction time with the levitated nanosphere is much shorter than its oscillation period. Examples include collisions with gas molecules~\cite{barker_collision-resolved_2024}, abrupt changes in mass or charge~\cite{ricci_accurate_2019}, or the emission of $\alpha$ particles~\cite{wang_mechanical_2024}.  According to the standard quantum limit (SQL) for impulsive force detection, the minimum resolvable momentum in absence of any quantum correlations is ${\rm Min}[\Delta p] = \sqrt{\hbar \:\!m\:\! \Omega_z}$. For a nanoparticle with mass $m=1\,$fg and a trap frequency 
of  $\Omega_z= 2\pi\times 60\,$kHz this  evaluates to ${\rm Min}[\Delta p] =  
11.8\,$keV/c~\cite{clerk_quantum-limited_2004}, a value that is more than three orders of magnitude smaller than the recoil of an $\alpha$ particle~\cite{wang_mechanical_2024} and is comparable to a thermal gas molecule. Thus, in principle, a levitated nanoparticle should be able to resolve collisions with single atoms~\cite{chaste12, yang11,cronin09,grier09,zipkes10}. 

In this work, we generate a time-gated, directional stream of atoms with well-defined velocity distribution, and direct it toward a levitated nanosphere cooled to its motional quantum ground state. We measure the displacement induced by the momentum transfer during individual collisions, with a sensitivity limited by the zero-point fluctuations of the nanosphere's translational motion.  
}

\paragraph{Experimental set-up.}
\begin{figure*}[t]
\centering
\includegraphics[width=\textwidth]{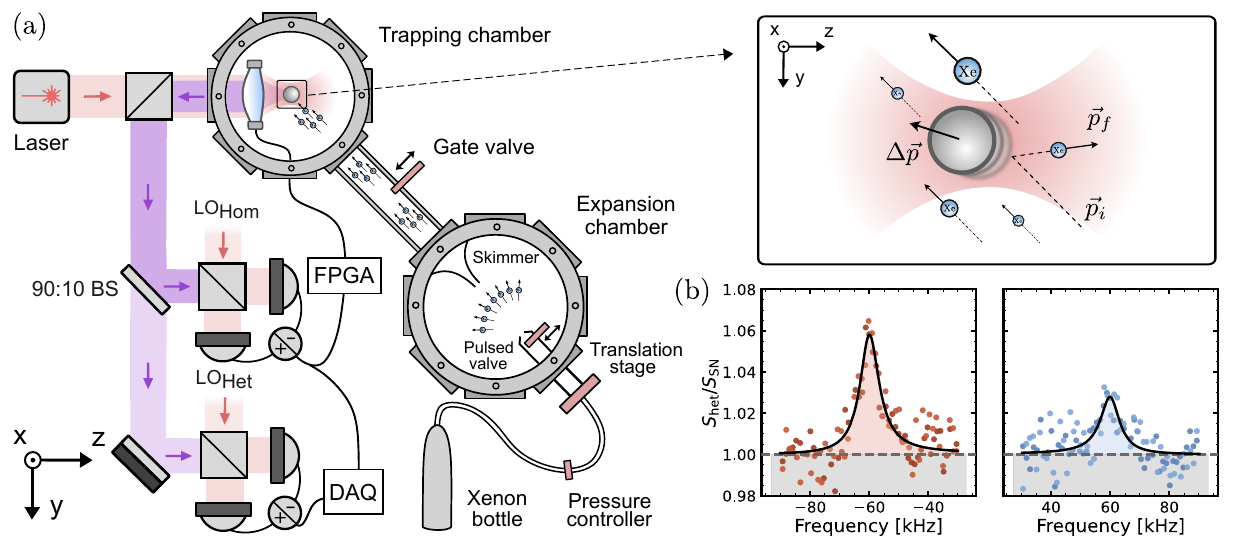}
\caption{\label{fig:experimental_set_up} 
(a) A silica nanosphere is optically trapped in high vacuum within the trapping chamber (top left). Its motion is detected via interferometric readout, in which  backscattered light is combined with a local oscillator and measured with a balanced photodetector (BPD). A supersonic beam of Xe atoms is produced by a pulsed valve in the expansion chamber (bottom right). The atomic beam is subsequently collimated by a skimmer and directed toward the levitated particle through a connecting tube between the two chambers. The close-up on the top right illustrates collisions between the nanosphere and single Xe atoms. (b) Stokes and anti-Stokes sidebands of the heterodyne spectrum from which we extract a phonon occupation number of $n = 0.80 \pm 0.30$. $S_{\rm het}$ and $S_{\rm SN}$ denote the power spectral densities of the heterodyne signal and the measurement imprecision noise, respectively.}
\end{figure*}

A silica nanosphere (diameter $100~\mathrm{nm}$) is trapped in an optical tweezer 
(power $0.7~\mathrm{W}$, wavelength $1550~\mathrm{nm}$) focused with a high numerical-aperture lens ($\mathrm{NA} = 0.8$). The experimental setup is illustrated in Fig.~\ref{fig:experimental_set_up}(a). The levitated particle is trapped inside a vacuum chamber, referred to as the trapping chamber (top left in Fig.~\ref{fig:experimental_set_up}(a)), operated under high-vacuum conditions at a pressure of $P = 9.0 \times 10^{-9}~\mathrm{mbar}$. In this regime, the particle is driven by photon-recoil noise of the trapping laser~\cite{jain16}. The center-of-mass motion of the particle is measured by mixing $90\%$ of the backscattered light with a local oscillator in a homodyne detection scheme using a balanced photodetector.
The remaining $10\%$ of the backscattered light is used in a heterodyne detection scheme with a fiber-coupled balanced photodetector
to measure the phonon occupation. The transverse center-of-mass modes, with frequencies $(\Omega_x, \Omega_y)/2\pi = (166, 199)~\mathrm{kHz}$, are cooled via  parametric feedback~\cite{gieseler12}, while the axial $z$ mode, with frequency $\Omega_z/2\pi = 59.9~\mathrm{kHz}$, is cooled to a phonon occupation of $n_\text{fb} = 0.80 \pm 0.30$ using  cold damping~\cite{Tebbenjohanns2020}. 
This phonon occupation is extracted from the sideband asymmetry 
shown in Fig.~\ref{fig:experimental_set_up}(b). 

In a second vacuum chamber, referred to as the expansion chamber (bottom right in Fig.~\ref{fig:experimental_set_up}(a)), a pulsed supersonic beam of Xe is prepared before being directed toward the levitated nanosphere. A Xe gas bottle is connected to a pressure controller, allowing the pressure of the injected gas
to be regulated to approximately $1.2~\mathrm{bar}$. The gas enters a pulsed valve, equipped with a nozzle of $100~\mu\mathrm{m}$ diameter and operated with an opening time of $200~\mu\mathrm{s}$ and a repetition period of $200~\mathrm{ms}$,  corresponding to a duty cycle of $0.1\%$. 
The pressure difference between the pulsed valve ($1.2~\mathrm{bar}$) and the expansion chamber ($9.8\times 10^{-6}~\mathrm{mbar}$) forces the Xe atoms to undergo supersonic expansion. After few centimeters of free expansion, the atoms reach a terminal velocity of  $v_{\rm t}\sim 310~\mathrm{m/s}$~\cite{hogan2011deceleration}. A fraction of the atoms is then spatially filtered by a $100~\mu\mathrm{m}$ diameter skimmer and directed toward the levitated nanosphere through a tube connecting the two chambers. The diameter of the atomic beam at the location of the nanosphere is estimated to be a few millimeters. 
A gate valve placed between the two chambers can be used to block the atomic beam. 

{\color{black} 
\begin{figure}[!b]
\hspace{-2em}\includegraphics[width=0.9\columnwidth]{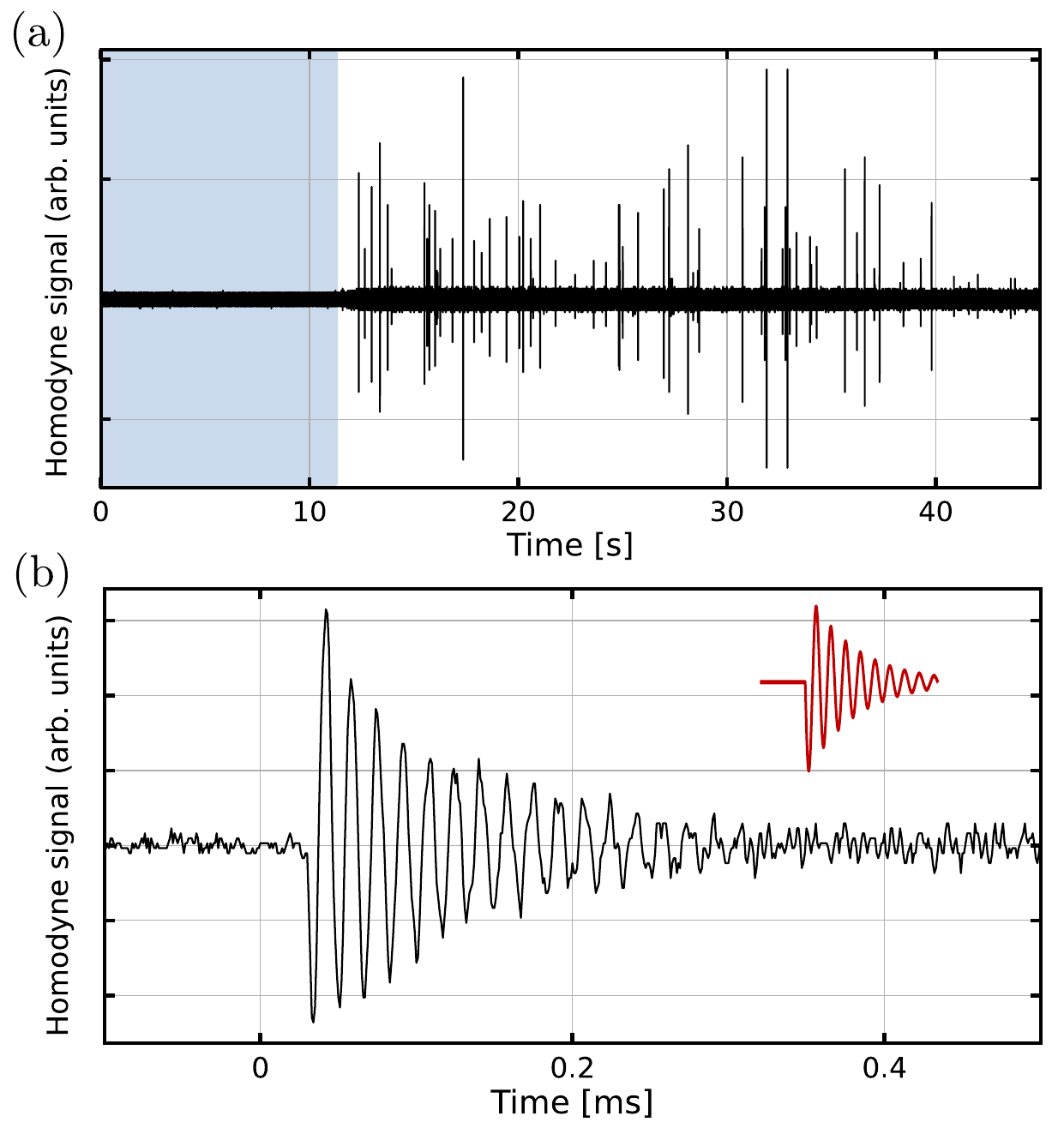}
\caption{\label{fig:cluster} 
{\color{black} Detection of Xe clusters. (a)~Detector time-trace showing a sequence of collision events. The valve is opened at $t\sim 11\,$s and remains open for later times. (b)~Zoom-in on the collision event at $t=13.74\,$s. The signal is in good agreement with the theoretical impulse response function (inset).}}
\end{figure}
To characterize the system's response function we first  expose the nanoparticle to a stream of Xe clusters, which are formed when the pressure difference between the injected gas and the expansion chamber exceeds a limit that depends on the nozzle diameter and other system parameters. The recoil imparted on the nanoparticle by a Xe cluster is substantial and hence the signal-to-noise ratio (SNR) of the measurement is large. Fig.~\ref{fig:cluster}(a) shows a sequence of detection events after opening the valve at $t\sim 11\,$s. In Fig.~\ref{fig:cluster}(b) we zoom in on a single detection event (at $t= 13.74\,$s) and find good agreement with the theoretical system response function (inset in Fig.~\ref{fig:cluster}b). The latter corresponds to the Green function of a damped harmonic oscillator, which reads
\begin{align}
h(t) \;=\; \frac{1}{\mathcal{N}} \: \Theta(t)\, {\rm e}^{-\gamma_{\rm eff} t/2}
\,\sin\left(\Omega_d t\right)\, .
\label{sigh}
\end{align}
$\Theta$ is the Heaviside function, $\mathcal{N}$ is a calibration factor, and $\gamma_{\rm eff}/2\pi=5.9~{\rm kHz}$ and $\Omega_d/2\pi = \sqrt{\Omega_z^2 - (\gamma_{\rm eff}/2)^2}/2\pi$ are the damping and trap frequency under feedback, respectively.  \\[-1ex]
}

\begin{figure*}[t]
\centering
\includegraphics[width=\textwidth]{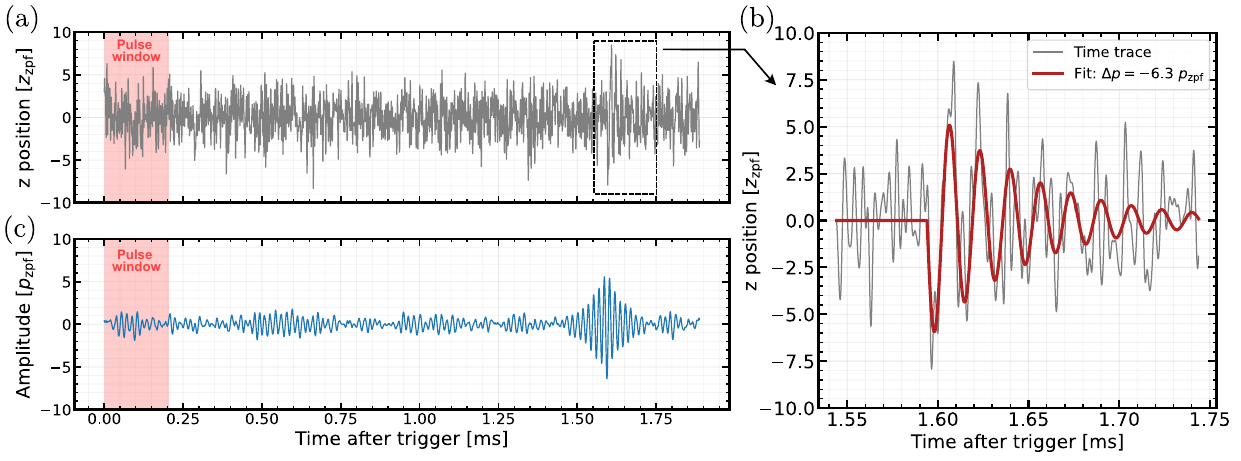}
\caption{\label{fig:collision} 
(a)~Measured homodyne position time-trace $z_m(t)$ recorded after triggering the opening of the pulsed valve at $t=0$. The time-trace is pre-filtered using a fourth-order high-pass filter at $5~\mathrm{kHz}$ and a fourth-order low-pass filter at $300~\mathrm{kHz}$. The $200~\mu\mathrm{s}$ time interval during which the pulsed valve is open is highlighted in red and referred to as the pulse window. 
{\color{black}{
(b)~Zoom-in on an impulsive event transferring $-6.3~p_{\mathrm{zpf}}$ of momentum to the levitated nanosphere (gray trace), along with the matched filter function of Eq.(\ref{sigh}) (red curve). (c)~Output of the matched filter,  corresponding to the convolution of the matched filter function with the recorded  time-trace $z_m(t)$.
}}}
\end{figure*}
\paragraph{Sensitivity to single collision events.}
{\color{black}{
We now turn to the detection of {\em single} Xe atoms.}} The SNR 
can be expressed as  ~\cite{tseng2025search,skrabulis2026nanomechanical} 
\begin{align}
    {\rm SNR}=|\Delta p \;\!/ \:\!\sigma_p |,
    \label{snreq}
\end{align}
where $\Delta p$ is the momentum transferred to the nanosphere and $\sigma_p$ is the conditional standard deviation of the nanosphere's momentum~\cite{rossi2019observing}. An upper bound of $\sigma_p$ is ${p_\mathrm{zpf}\sqrt{2\!\:n_\text{fb}\!+\!1}}$, with $p_\mathrm{zpf}$ denoting the zero-point fluctuations of the nanosphere's momentum and $n_\text{fb}$ the feedback-cooled occupation, which, at best, equals the conditional state variance~\cite{tebbenjohanns2021quantum}.

In our experiment, Xe atoms are incident at an angle of $45^{\circ}$ to the $z$ axis, resulting in a maximum momentum transfer along the $-z$ direction of $\Delta p_{\rm max} = (1\!+\!1/\sqrt{2})\:\! m_{\rm Xe} \:\!v_{\rm t} \sim 217\,$keV/c, where $m_{\rm Xe}=132~{\rm amu}$ is the Xe atomic mass. 
In units of zero-point fluctuations $p_{\rm zpf} =\sqrt{\hbar\:\!m \Omega_z/2}=8.54\,$keV/c, the maximum momentum transfer corresponds to $\Delta p_{\rm max} \sim 25 \,p_{\rm zpf}$. 
According to Eq.~(\ref{snreq}) we expect a ${\rm SNR}\sim 16$ for the strongest momentum kicks.

\paragraph{Results.} 
{\color{black} The pulsed valve allows us to generate well-separated collision events and to perform time-gated measurements.} In Fig.~\ref{fig:collision}(a) we show the position $z_{\rm m}(t)$ of the levitated particle, measured with the homodyne detector, as a function of time after the trigger, where $t=0$ corresponds to the opening of the pulsed valve. The measured signal, denoted $z_{\rm m}(t)$, corresponds to the sum of  $z(t)$  and the measurement imprecision noise (shot noise). 
The red shaded area
indicates the time interval during which the pulsed valve remains open. At a time of $t\sim 1.6~{\rm ms}$ (highlighted by the rectangle in Fig.~\ref{fig:collision}(a)) we observe a signal that is stronger than the noise floor. This event is magnified and depicted as the gray curve in Fig.~\ref{fig:collision}(b).
%
To identify such impulsive events, we apply a matched filter to the measured time-trace.  A matched filter is the optimal filter for estimating the amplitude and arrival time of a known signal in the presence of additive stochastic noise~\cite{kay1998}. In our case, the signal corresponds to the nanosphere's displacement due to an impulsive excitation, that is, to Eq.~(\ref{sigh}).
{\color{black}{For the feedback gains used in our experiments, the background noise is, to a good approximation, Gaussian white noise~\cite{tebbenjohanns2021quantum}.}} Under these conditions, matched filtering corresponds to cross-correlating the measured signal $z_{\rm m}(t)$ with the template $h(t)$. In Fig.~\ref{fig:collision}(c), we plot the filtered position time-trace in units of $p_{\rm zpf}$ and observe, as expected, the largest signal at $1.6~{\rm ms}$ after the trigger,  corresponding to a momentum kick of $-6.3~p_{\rm zpf}$. 
 For every momentum kick, the matched filter renders both the collision time and the magnitude of the nanosphere's recoil.  In Fig.~\ref{fig:collision}(b), we overlay the inferred signal (red curve) to the raw signal at $1.6~{\rm ms}$.

Next, we perform a statistical analysis over multiple realizations of the experiment by acquiring $9,958$ time-traces, each $10~{\rm ms}$ long, starting after the valve trigger. We apply  matched filtering to each of these time-traces to identify single collisions. For each time-trace, we first identify the largest event and record the corresponding collision time $t_0$ and momentum $\Delta p_0$.  We exclude a time window of $\Delta t = \pm~0.2~{\rm ms}$ centered at the peak of this event to avoid double-counting. We repeat this procedure iteratively until no events above the detection threshold of $4 \:\!p_{\rm zpf}$ remain. The choice for this threshold and for the duration of the time window are discussed later in the text.

\begin{figure}[!b]
\includegraphics[width=0.9\columnwidth]{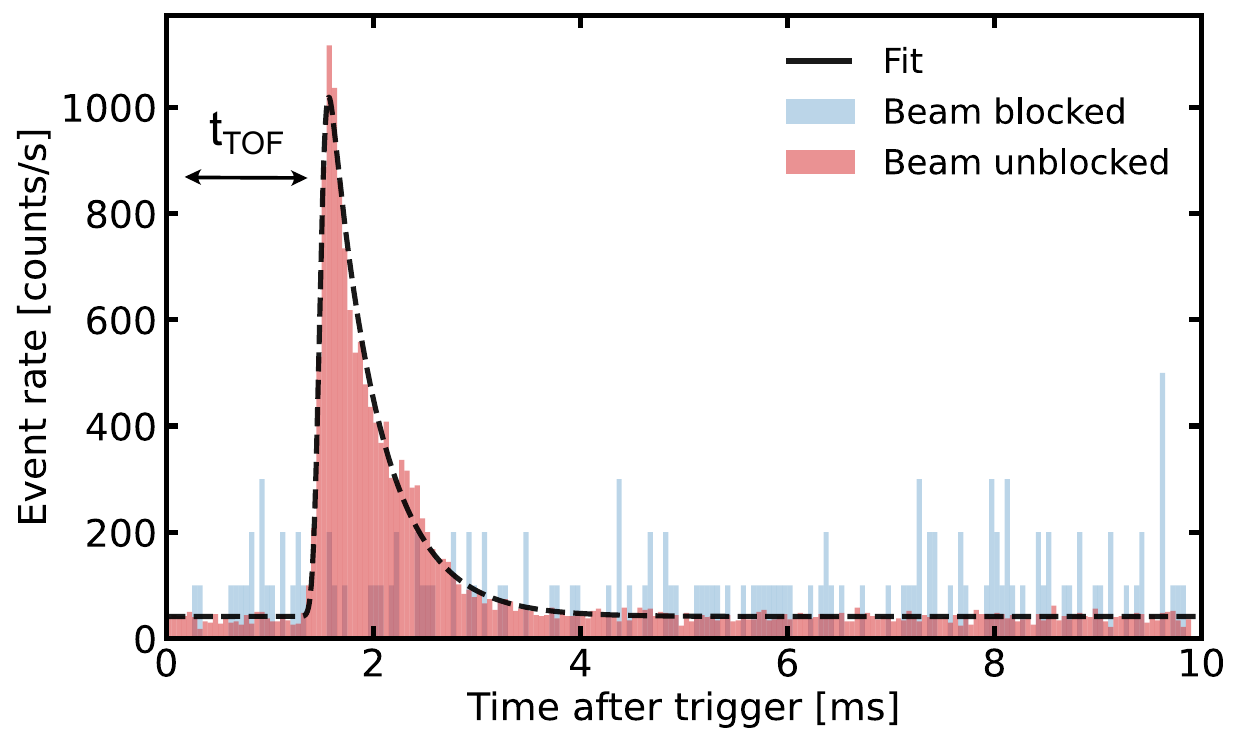}
\caption{\label{fig:time_and_position} 
Event rate (counts per second) above a $4\:\!p_{\mathrm{zpf}}$ threshold as a function of time after opening the pulsed valve. The red histogram is recorded with the atomic beam unblocked (average over $9,958$ time-traces). The blue histogram is acquired with the beam blocked (average over $200$ time-traces). 
}
\end{figure}

In Fig.~\ref{fig:time_and_position}, we plot the event rate (number of detected events per second) as a function of the time after the trigger (red histogram), using a bin width of $50~\mu{\rm s}$. We observe that the event rate remains flat at  $\sim 40~{\rm counts/s}$ for approximately $1.5~{\rm ms}$ after the opening of the pulsed valve. We then observe a sharp increase in the event rate, reaching values above $1000~{\rm counts/s}$. The event rate then decreases over a few milliseconds, following an approximately exponential decay. After $5~{\rm ms}$, the event rate is similar to the one observed during the first $1.5~{\rm ms}$ after opening of the valve. 
We fit the rising edge of the histogram to an erf function, which yields a time of flight of $t_\text{TOF}=1.47~$ms. This time, together with the distance $44\pm2$~cm between the pulsed valve and the nanosphere, yields an atomic velocity of $299\pm14$~m/s, in good agreement with the terminal velocity of supersonic Xe ($\approx 310~$m/s)~\cite{hogan2011deceleration}.
Furthermore, we fit the tail of the histogram to an exponential decay (decay time $\tau=0.47~$ms), which we attribute to plunger-rebounces in the pulsed valve. These rebounces give rise to weaker secondary gas pulses immediately following the main pulse. Such effects are common in our type of pulsed valve (see Fig.~7 in~\cite{allmendinger2016new}).
The observed tail can also have contributions from Xe atoms clogging the skimmer, and consequently losing velocity, an effect expected for a $100~\mu{\rm m}$-diameter skimmer~\cite{bird1976transition,even2014pulsed}.

%

As a control experiment, we determine the event rate with the gate valve between the two vacuum chambers partially closed [blue histogram in Fig.~\ref{fig:time_and_position}].
Partially closing the gate valve blocks the supersonic beam, while still allowing Xe atoms to diffuse into the trapping chamber. This results in a pressure increase that is similar to the case of the unblocked beam. We observe that, when the beam is blocked, the event rate remains flat at a value close to $40~{\rm counts/s}$ (the same value observed with the atomic beam unobstructed) during the whole $10~{\rm ms}$. 
Therefore, the flat background observed in the measurement with the atomic beam unobstructed (red data in Fig.~\ref{fig:time_and_position}) can be attributed to collisions of the nanosphere with thermal atoms, while the sharp spike stems from collisions with Xe atoms forming the beam.
Note that, for the control dataset, the number of acquisitions is smaller ($200$ time-traces), which explains the larger variance.

To provide further evidence that the background rate of 40~counts/s is associated with thermal Xe atoms, we close the gate valve entirely and purge the chamber with nitrogen gas to ensure that practically no Xe remains in the trapping chamber. We observe that the event rate drops to 3.6 counts/s  (data not shown), which corresponds to the ``dark count rate'' of our detector that was used to determine the threshold value of $4\:\!p_\text{zpf}$ throughout this work. Under this choice, 99\% of the events detected at the peak of the red histogram in Fig.~\ref{fig:time_and_position} can be attributed to collisions with Xe atoms.

As a technical aside, a single event could, in principle, be counted twice if the matched filter output remained above the threshold value of $4\:\!p_{\rm zpf}$ outside the $\Delta t=\pm 0.2~{\rm ms}$ window. However, this does not occur in our measurements, since all detected events remain below $20\:\!p_{\rm zpf}$. Taking into account the characteristic matched filter decay, $\mathrm{e}^{-\gamma_{\rm eff} |\Delta t|/2} \approx 3\%$, the residual signal amplitude outside the time window remains well below the detection threshold of $4\:\!p_{\rm zpf}$.

\begin{figure}[!t]
\includegraphics[width=0.99\columnwidth]{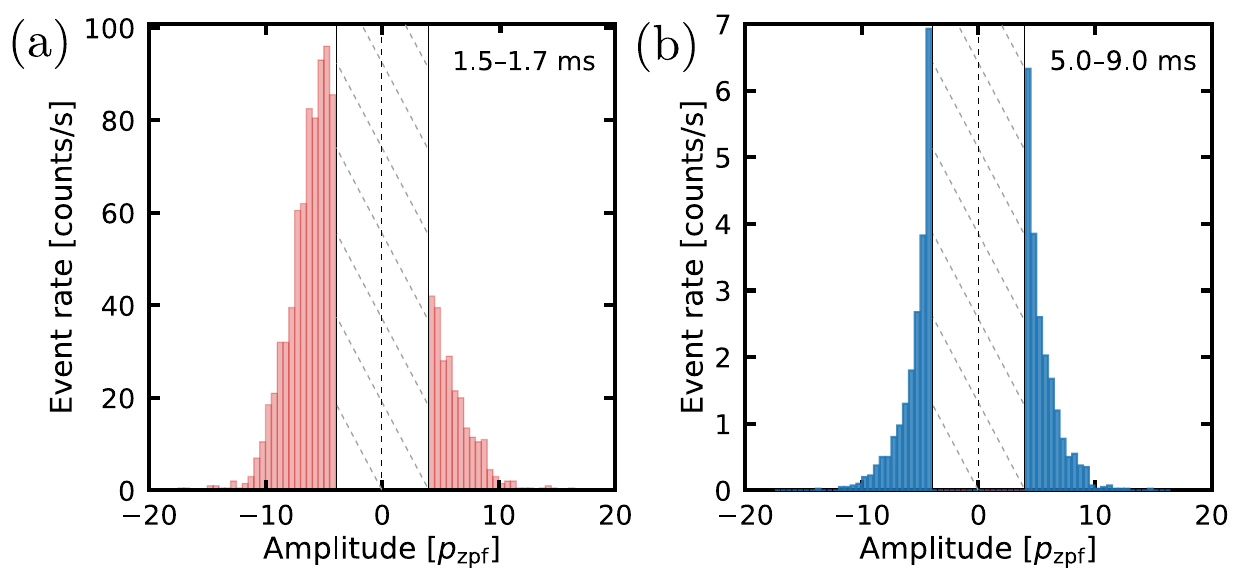}
\caption{\label{fig:probability_distribution} 
{\color{black}{Histogram of collision amplitudes evaluated for two different time periods after valve opening. (a) for events detected within $1.5$–$1.7~{\rm ms}$ (directional atoms) and (b) for events detected within  $5.0$–$9.0~{\rm ms}$ (thermal atoms). The central region of $[-4\, .. \,4]\,p_{zpf}$ is below the detection threshold and is blanked out.}}}
\end{figure}

We continue by discussing the directionality of the observed momentum transfer. 
Following a collision event, an initially negative displacement of the measured position indicates a displacement along the $-z$ direction, 
such as the event observed in Fig.~\ref{fig:collision}(b). {\color{black}{We refer to momentum kicks in $-z$ direction as forward-scattering events, while  momentum kicks in $+z$ direction are denoted backward-scattering events.}} In Fig.~\ref{fig:probability_distribution}, we plot the event rate as a function of the measured amplitude for two different time windows after the trigger: $1.5$--$1.7~{\rm ms}$ (red), associated with supersonic Xe atoms, and $5.0$--$9.0~{\rm ms}$ (blue), associated with thermal Xe atoms. As expected, the histogram 
for thermal Xe atoms is symmetric, indicating that forward- and backward-scattering events are equally probable. In contrast, the histogram associated with supersonic Xe atoms is asymmetric, exhibiting a larger number of forward-scattering events. {\color{black}{The residual backward-scattering events have different origins, such as contributions from thermalized atoms and from atoms striking the surface of the nanosphere at shallow angles (recall the $45^{\circ}$ alignment between the atomic beam and the measurement axis $z$). 
Also, the Xe beam size is comparable to critical geometrical dimensions (e.g.  focal length of trapping lens), giving rise to spurious reflections and secondary scattering events.
These experimental constraints prevent us currently from establishing a quantitative comparison with a  theoretical model. Nevertheless, the symmetries of the distributions in Fig.~\ref{fig:probability_distribution} clearly capture the difference between a thermal gas and a directional beam of atoms.}}
%
{\color{black}{
From the event rate distribution in Fig.~\ref{fig:probability_distribution}(a) it follows that 
the event rate for $-5\, p_{zpf} = 42.7\,$keV/c  is approximately $90\,$counts/s. On the other hand, according to Fig.~\ref{fig:probability_distribution}(b), the background rate for the same event rate is $3.6\,$counts/s. The ratio between the two indicates that we 
can measure collisions below $50~{\rm keV}/c$ with a confidence level of $(90-3.6)/90 = 96\%$.\\[-1ex]}}

\paragraph{Conclusion.}
We have studied collisions between individual Xe atoms and a levitated nanosphere cooled to its motional quantum ground state.
Using supersonic expansion in combination with a pulsed valve, we have achieved control over the collision times, as well as over the direction and velocity of the incident atoms. Our experimental results demonstrate the detection of impulsive forces smaller than $50~{\rm keV}/c$ with $96\%$ confidence. This sensitivity opens the door for studying particles in unexplored parameter regimes~\cite{moore2021searching,tseng2025search} and for collision-resolved pressure sensing~\cite{barker_collision-resolved_2024}. The implementation of squeezing and coherent amplification will further enhance the sensitivity~\cite{skrabulis2026nanomechanical} and enable the detection of even lighter atoms and molecules.

In our experiments we ignored the internal states of the atoms and only considered the center-of-mass degrees of freedom. Exploring interaction protocols in which the internal degrees of freedom can be harnessed (e.g., via coupling to cavity modes) would be of interest for the generation of nonclassical states of the levitated nanosphere. Furthermore, the nanosphere can serve as a detector in neutral-atom interferometry experiments, eliminating the need for ionization or preselection of metastable species~\cite{PhysRevLett.66.2689}. \\

\paragraph{Note added.} We recently became aware of a related independent work by Tseng et al.~\cite{tseng2026optomechanical}.\\

\subsection{Acknowledgments}
We thank Loris Coccia for his assistance during the early stages of this work. We thank Maria Luisa Mattana and Martynas Skrabulis for sharing their knowledge on ground-state cooling of levitated nanospheres, and the rest of our colleagues at the ETH Photonics Laboratory for fruitful discussions. This research has been supported by NCCR Precision, a National Centre of Competence in Research funded by the Swiss National Science Foundation (Grant 51AU-0\_229299) and the Swiss SERI Quantum Initiative (Grants No. UeM019-2 and No. UeM029-3).\\


\bibliography{references.bib}

@article{jain16,
  title = {Direct Measurement of Photon Recoil from a Levitated Nanoparticle},
  author = {Jain, Vijay and Gieseler, Jan and Moritz, Clemens and Dellago, Christoph and Quidant, Romain and Novotny, Lukas},
  journal = {Phys. Rev. Lett.},
  volume = {116},
  issue = {24},
  pages = {243601},
  numpages = {5},
  year = {2016},
  month = {Jun},
  publisher = {American Physical Society},
  doi = {10.1103/PhysRevLett.116.243601},
  url = {https://link.aps.org/doi/10.1103/PhysRevLett.116.243601}
}

@misc{jin_towards_2024,
  title = {Towards Real-World Applications of Levitated Optomechanics},
  author = {Jin, Yuanbin and Shen, Kunhong and Ju, Peng and Li, Tongcang},
  year = 2024,
  month = jul,
  number = {arXiv:2407.12496},
  eprint = {2407.12496},
  primaryclass = {physics},
  publisher = {arXiv},
  doi = {10.48550/arXiv.2407.12496},
  urldate = {2025-05-23},
  archiveprefix = {arXiv}
}

@article{clerk_quantum-limited_2004,
  title = {Quantum-Limited Position Detection and Amplification: {{A}} Linear Response Perspective},
  shorttitle = {Quantum-Limited Position Detection and Amplification},
  author = {Clerk, A. A.},
  year = 2004,
  month = dec,
  journal = {Phys. Rev. B},
  volume = {70},
  number = {24},
  pages = {245306},
  issn = {1098-0121, 1550-235X},
  doi = {10.1103/PhysRevB.70.245306},
  urldate = {2020-06-18},
  langid = {english}
}

@article{gieseler12,
  title={Probing quantum mechanics with nanoparticle matter-wave interferometry},
  author={Pedalino, Sebastian and Ram{\'\i}rez-Galindo, Bruno E and Ferstl, Richard and Hornberger, Klaus and Arndt, Markus and Gerlich, Stefan},
  journal={Nature},
  url={https://doi.org/10.1038/s41586-025-09917-9},
  volume={649},
  number={8098},
  pages={866--870},
  year={2026},
  publisher={Nature Publishing Group UK London}
}

@article{gonzalez2021levitodynamics,
  title={Levitodynamics: Levitation and control of microscopic objects in vacuum},
  author={Gonzalez-Ballestero, Carlos and Aspelmeyer, Markus and Novotny, Lukas and Quidant, Romain and Romero-Isart, Oriol},
  journal={Science},
  url={https://www.science.org/doi/10.1126/science.abg3027},
  volume={374},
  pages = {eabg3027},
  number={6564},
  year={2021},
  publisher={American Association for the Advancement of Science}
}

@article{frimmer_controlling_2017,
  title = {Controlling the Net Charge on a Nanoparticle Optically Levitated in Vacuum},
  author = {Frimmer, Martin and Luszcz, Karol and Ferreiro, Sandra and Jain, Vijay and Hebestreit, Erik and Novotny, Lukas},
  year = 2017,
  month = jun,
  journal = {Phys. Rev. A},
  volume = {95},
  number = {6},
  pages = {061801},
  issn = {2469-9926, 2469-9934},
  doi = {10.1103/PhysRevA.95.061801},
  urldate = {2021-12-06},
  langid = {english}
}

@article{tebbenjohanns2021quantum,
  title={Quantum control of a nanoparticle optically levitated in cryogenic free space},
  author={Tebbenjohanns, Felix and Mattana, M Luisa and Rossi, Massimiliano and Frimmer, Martin and Novotny, Lukas},
  url={https://doi.org/10.1038/s41586-021-03617-w},
  journal={Nature},
  volume={595},
  number={7867},
  pages={378--382},
  year={2021},
  publisher={Nature Publishing Group UK London}
}

@article{rossi2019observing,
  title={Observing and verifying the quantum trajectory of a mechanical resonator},
  author={Rossi, Massimiliano and Mason, David and Chen, Junxin and Schliesser, Albert},
  journal={Physical Review Letters},
  url = {https://link.aps.org/doi/10.1103/PhysRevLett.123.163601},
  volume={123},
  number={16},
  pages={163601},
  year={2019},
  publisher={APS}
}

@article{PhysRevLett.66.2689,
  title = {Young's double-slit experiment with atoms: A simple atom interferometer},
  author = {Carnal, O. and Mlynek, J.},
  journal = {Phys. Rev. Lett.},
  volume = {66},
  issue = {21},
  pages = {2689--2692},
  numpages = {0},
  year = {1991},
  month = {May},
  publisher = {American Physical Society},
  doi = {10.1103/PhysRevLett.66.2689},
  url = {https://link.aps.org/doi/10.1103/PhysRevLett.66.2689}
}

@article{Tebbenjohanns2020,
  title = {Motional Sideband Asymmetry of a Nanoparticle Optically Levitated in Free Space},
  author = {Tebbenjohanns, Felix and Frimmer, Martin and Jain, Vijay and Windey, Dominik and Novotny, Lukas},
  journal = {Phys. Rev. Lett.},
  volume = {124},
  issue = {1},
  pages = {013603},
  numpages = {4},
  year = {2020},
  month = {Jan},
  publisher = {American Physical Society},
  doi = {10.1103/PhysRevLett.124.013603},
  url = {https://link.aps.org/doi/10.1103/PhysRevLett.124.013603}
}

@article{allmendinger2016new,
  title={New method to study ion--molecule reactions at low temperatures and application to the reaction},
  author={Allmendinger, Pitt and Deiglmayr, Johannes and Schullian, Otto and H{\"o}veler, Katharina and Agner, Josef A and Schmutz, Hansj{\"u}rg and Merkt, Fr{\'e}d{\'e}ric},
  journal={ChemPhysChem},
  volume={17},
  number={22},
  pages={3596--3608},
  year={2016},
  publisher={Wiley Online Library}
}

@article{romero-isart_quantum_2011,
  title = {Quantum Superposition of Massive Objects and Collapse Models},
  author = {{Romero-Isart}, Oriol},
  year = 2011,
  month = nov,
  journal = {Phys. Rev. A},
  volume = {84},
  number = {5},
  pages = {052121},
  issn = {1050-2947, 1094-1622},
  doi = {10.1103/PhysRevA.84.052121},
  urldate = {2020-07-24},
  langid = {english}
}

@book{kay1998,
  title={Fundamentals of Statistical Signal Processing: Detection Theory},
  author={Steven M. Kay},
  year={1998},
  publisher={Prentice-Hall PTR}
}

@article{bird1976transition,
  title={Transition regime behavior of supersonic beam skimmers},
  author={Bird, GA},
  journal={The Physics of Fluids},
  url={https://doi.org/10.1063/1.861351},
  volume={19},
  number={10},
  pages={1486--1491},
  year={1976},
  publisher={American Institute of Physics}
}

@article{even2014pulsed,
  title={Pulsed supersonic beams from high pressure source: Simulation results and experimental measurements},
  author={Even, Uzi},
  journal={Advances in Chemistry},
  volume={2014},
  number={1},
  pages={636042},
  year={2014},
  publisher={Wiley Online Library}
}

@article{skrabulis2026nanomechanical,
  title={Nanomechanical sensor resolving impulsive forces below its zero-point fluctuations},
  author={Skrabulis, Martynas and Sosa, Martin Colombano and Zambon, Nicola Carlon and Militaru, Andrei and Rossi, Massimiliano and Frimmer, Martin and Novotny, Lukas},
  journal={Phys. Rev. Lett.},
  volume={136},
  pages={233604},
  year={2026},
  publisher={APS}
}

@article{tseng2025search,
  title={Search for dark matter scattering from optically levitated nanoparticles},
  author={Tseng, Yu-Han and Penny, TW and Siegel, Benjamin and Wang, Jiaxiang and Moore, David C},
  journal={PRX Quantum},
  volume={6},
  number={4},
  pages={040367},
  year={2025},
  publisher={APS}
}

@article{hempston_force_2017,
  title = {Force Sensing with an Optically Levitated Charged Nanoparticle},
  author = {Hempston, David and Vovrosh, Jamie and Toro{\v s}, Marko and Winstone, George and Rashid, Muddassar and Ulbricht, Hendrik},
  year = 2017,
  month = sep,
  journal = {Applied Physics Letters},
  volume = {111},
  number = {13},
  publisher = {AIP Publishing},
  issn = {0003-6951, 1077-3118},
  doi = {10.1063/1.4993555},
  urldate = {2025-05-23},
  langid = {english}
}

@article{moore2021searching,
  title={Searching for new physics using optically levitated sensors},
  author={Moore, David C and Geraci, Andrew A},
  journal={Quantum Science \& Technology},
  url = {https://doi.org/10.1088/2058-9565/abcf8a},
  volume={6},
  number={1},
  pages={014008},
  year={2021},
  publisher={IOP Publishing}
}

@article{shi_mobile_2025,
  title = {A {{Mobile Electric Field Sensor Device Based}} on {{Optically Levitated Nano-resonators}}},
  author = {Shi, Yunjie and Chen, Zhiming and Zhang, Yizhou and Wang, Yuehao and He, Peitong and Zhu, Xunmin and Zheng, Yi and Wang, Yingying and Guo, Leilei and Wu, Bin and Fu, Zhenhai and Gao, Xiaowen and Hu, Huizhu},
  year = 2025,
  month = jan,
  journal = {Adv Devices Instrum},
  volume = {6},
  publisher = {American Association for the Advancement of Science (AAAS)},
  issn = {2767-9713},
  doi = {10.34133/adi.0090},
  urldate = {2025-05-23},
}

@article{fuchs_measuring_2024,
  title = {Measuring Gravity with Milligram Levitated Masses},
  author = {Fuchs, Tim M. and Uitenbroek, Dennis G. and Plugge, Jaimy and Van Halteren, Noud and Van Soest, Jean-Paul and Vinante, Andrea and Ulbricht, Hendrik and Oosterkamp, Tjerk H.},
  year = 2024,
  month = feb,
  journal = {Sci. Adv.},
  volume = {10},
  number = {8},
  pages = {eadk2949},
  issn = {2375-2548},
  doi = {10.1126/sciadv.adk2949},
  urldate = {2024-11-11},
  langid = {english}
}

@article{moore_search_2014,
  title = {Search for {{Millicharged Particles Using Optically Levitated Microspheres}}},
  author = {Moore, David C. and Rider, Alexander D. and Gratta, Giorgio},
  year = 2014,
  month = dec,
  journal = {Phys. Rev. Lett.},
  volume = {113},
  number = {25},
  pages = {251801},
  issn = {0031-9007, 1079-7114},
  doi = {10.1103/PhysRevLett.113.251801},
  urldate = {2020-07-30},
  langid = {english}
}

@article{ricci_accurate_2019,
  title = {Accurate {{Mass Measurement}} of a {{Levitated Nanomechanical Resonator}} for {{Precision Force-Sensing}}},
  author = {Ricci, F. and Cuairan, M. T. and Conangla, G. P. and Schell, A. W. and Quidant, R.},
  year = 2019,
  month = oct,
  journal = {Nano Lett.},
  volume = {19},
  number = {10},
  pages = {6711--6715},
  publisher = {American Chemical Society (ACS)},
  issn = {1530-6984, 1530-6992},
  doi = {10.1021/acs.nanolett.9b00082},
  urldate = {2025-05-23},
  copyright = {https://doi.org/10.15223/policy-029},
  langid = {english}
}

@article{hebestreit_sensing_2018,
  title = {Sensing {{Static Forces}} with {{Free-Falling Nanoparticles}}},
  author = {Hebestreit, Erik and Frimmer, Martin and Reimann, Ren{\'e} and Novotny, Lukas},
  year = 2018,
  month = aug,
  journal = {Phys. Rev. Lett.},
  volume = {121},
  number = {6},
  pages = {063602},
  issn = {0031-9007, 1079-7114},
  doi = {10.1103/PhysRevLett.121.063602},
  urldate = {2020-08-08},
  langid = {english}
}

@article{liu_nanoscale_2024,
  title = {Nanoscale Vacuum Gauge Based on Second-Order Coherence in Optical Levitation},
  author = {Liu, Lyu-Hang and Zheng, Yu and Tian, Yuan and Wang, Long and Guo, Guang-Can and Sun, Fang-Wen},
  year = 2024,
  month = oct,
  journal = {Phys. Rev. Applied},
  volume = {22},
  number = {4},
  publisher = {American Physical Society (APS)},
  issn = {2331-7019},
  doi = {10.1103/physrevapplied.22.l041006},
  urldate = {2025-05-23},
  copyright = {https://link.aps.org/licenses/aps-default-license},
  langid = {english}
}

@article{barker_collision-resolved_2024,
  title = {Collision-Resolved Pressure Sensing},
  author = {Barker, Daniel S. and Carney, Daniel and LeBrun, Thomas W. and Moore, David C. and Taylor, Jacob M.},
  year = 2024,
  month = apr,
  journal = {Phys. Rev. A},
  volume = {109},
  number = {4},
  pages = {042616},
  issn = {2469-9926, 2469-9934},
  doi = {10.1103/PhysRevA.109.042616},
  urldate = {2024-06-07},
  langid = {english}
}

@article{ahrens_levitated_2025,
  title = {Levitated Ferromagnetic Magnetometer with Energy Resolution Well Below $\hbar$},
  author = {Ahrens, Felix and Ji, Wei and Budker, Dmitry and Timberlake, Chris and Ulbricht, Hendrik and Vinante, Andrea},
  year = 2025,
  month = mar,
  journal = {Phys. Rev. Lett.},
  volume = {134},
  number = {11},
  publisher = {American Physical Society (APS)},
  issn = {0031-9007, 1079-7114},
  doi = {10.1103/physrevlett.134.110801},
  urldate = {2025-05-23},
  copyright = {https://link.aps.org/licenses/aps-default-license},
  langid = {english}
}

@article{timberlake_acceleration_2019,
  title = {Acceleration Sensing with Magnetically Levitated Oscillators above a Superconductor},
  author = {Timberlake, Chris and Gasbarri, Giulio and Vinante, Andrea and Setter, Ashley and Ulbricht, Hendrik},
  year = 2019,
  month = nov,
  journal = {Applied Physics Letters},
  volume = {115},
  number = {22},
  publisher = {AIP Publishing},
  issn = {0003-6951, 1077-3118},
  doi = {10.1063/1.5129145},
  urldate = {2025-05-23}
}

@article{hogan2011deceleration,
  title={Deceleration of supersonic beams using inhomogeneous electric and magnetic fields},
  author={Hogan, Stephen D and Motsch, Michael and Merkt, Fr{\'e}d{\'e}ric},
  journal={Physical Chemistry Chemical Physics},
  url={https://doi.org/10.1039/c1cp21733j},
  volume={13},
  number={42},
  pages={18705--18723},
  year={2011},
  publisher={Royal Society of Chemistry}
}

@article{wang_mechanical_2024,
  title = {Mechanical {{Detection}} of {{Nuclear Decays}}},
  author = {Wang, Jiaxiang and Penny, T. W. and Recoaro, Juan and Siegel, Benjamin and Tseng, Yu-Han and Moore, David C.},
  year = 2024,
  month = jul,
  journal = {Phys. Rev. Lett.},
  volume = {133},
  number = {2},
  pages = {023602},
  issn = {0031-9007, 1079-7114},
  doi = {10.1103/PhysRevLett.133.023602},
  urldate = {2024-11-11},
  langid = {english}
}

@article{chaste12,
  title = {A nanomechanical mass sensor with yoctogram resolution},
  author = {J. Chaste and A. Eichler and J. Moser and G. Ceballos and R. Rurali and A. Bachtold},
  year = 2012,
  journal = {Nature Nanotech.},
  volume = {7},
  pages = {301--304}
}

@article{yang11,
author = {Yang, Y. T. and Callegari, C. and Feng, X. L. and Roukes, M. L.},
title = {Surface Adsorbate Fluctuations and Noise in Nanoelectromechanical Systems},
journal = {Nano Letters},
volume = {11},
number = {4},
pages = {1753-1759},
year = {2011},
doi = {10.1021/nl2003158},
    note ={PMID: 21388120},
URL = {https://doi.org/10.1021/nl2003158},
eprint = {https://doi.org/10.1021/nl2003158}
}

@article{grier09,
  title = {Observation of Cold Collisions between Trapped Ions and Trapped Atoms},
  author = {Grier, Andrew T. and Cetina, Marko and Oru\ifmmode \check{c}\else \v{c}\fi{}evi\ifmmode \acute{c}\else \'{c}\fi{}, Fedja and Vuleti\ifmmode \acute{c}\else \'{c}\fi{}, Vladan},
  journal = {Phys. Rev. Lett.},
  volume = {102},
  issue = {22},
  pages = {223201},
  numpages = {4},
  year = {2009},
  month = {Jun},
  publisher = {American Physical Society},
  doi = {10.1103/PhysRevLett.102.223201},
  url = {https://link.aps.org/doi/10.1103/PhysRevLett.102.223201}
}

@article{zipkes10,
  title = {A trapped single ion inside a Bose?Einstein condensate},
  author = {Christoph Zipkes and Stefan Palzer and Carlo Sias and Michael K\"ohl },
  journal = {Nature},
  volume = {464},
  pages = {388--391},
  year = {2010}
  }

@article{cronin09,
  title = {Optics and interferometry with atoms and molecules},
  author = {Cronin, Alexander D. and Schmiedmayer, J\"org and Pritchard, David E.},
  journal = {Rev. Mod. Phys.},
  volume = {81},
  issue = {3},
  pages = {1051--1129},
  numpages = {0},
  year = {2009},
  month = {Jul},
  publisher = {American Physical Society},
  doi = {10.1103/RevModPhys.81.1051},
  url = {https://link.aps.org/doi/10.1103/RevModPhys.81.1051}
}

@article{monteiro2020search,
  title = {Search for {{Composite Dark Matter}} with {{Optically Levitated Sensors}}},
  author = {Monteiro, Fernando and Afek, Gadi and Carney, Daniel and Krnjaic, Gordan and Wang, Jiaxiang and Moore, David C.},
  year = 2020,
  month = oct,
  journal = {Phys. Rev. Lett.},
  volume = {125},
  number = {18},
  pages = {181102},
  issn = {0031-9007, 1079-7114},
  doi = {10.1103/PhysRevLett.125.181102},
  urldate = {2024-09-18},
  langid = {english}
}

@article{zeng_optically_2024,
  title = {Optically Levitated Micro Gyroscopes with an {{MHz}} Rotational Vaterite Rotor},
  author = {Zeng, Kai and Xu, Xiangming and Wu, Yulie and Wu, Xuezhong and Xiao, Dingbang},
  year = 2024,
  month = jun,
  journal = {Microsyst Nanoeng},
  volume = {10},
  number = {1},
  pages = {78},
  issn = {2055-7434},
  doi = {10.1038/s41378-024-00726-0},
  urldate = {2025-01-14},
  langid = {english}
}

@article{zielinska_long-axis_2024,
  title = {Long-{{Axis Spinning}} of an {{Optically Levitated Particle}}: {{A Levitated Spinning Top}}},
  shorttitle = {Long-{{Axis Spinning}} of an {{Optically Levitated Particle}}},
  author = {Zieli{\'n}ska, J. A. and Van Der Laan, F. and Norrman, A. and Reimann, R. and Frimmer, M. and Novotny, L.},
  year = 2024,
  month = jun,
  journal = {Phys. Rev. Lett.},
  volume = {132},
  number = {25},
  publisher = {American Physical Society (APS)},
  issn = {0031-9007, 1079-7114},
  doi = {10.1103/physrevlett.132.253601},
  urldate = {2025-05-23},
  copyright = {https://link.aps.org/licenses/aps-default-license},
  langid = {english}
}

@article{tseng2026optomechanical,
  title={Optomechanical Detection of Individual Gas Collisions},
  author={Tseng, Yu-Han and Hardy, Clarke A and Penny, TW and Lowe, Cecily and Baeza-Rubio, Jacqueline and Carney, Daniel and Moore, David C},
  url={https://arxiv.org/abs/2604.18371},
  journal={arXiv preprint arXiv:2604.18371},
  year={2026}
}

\end{document}